\pdfoutput=1

\documentclass[11pt]{article}

\usepackage{amsmath}
\usepackage[final]{acl}

\usepackage{times}
\usepackage{latexsym}

\usepackage[T1]{fontenc}

\usepackage[utf8]{inputenc}

\usepackage{microtype}

\usepackage{inconsolata}

\usepackage{graphicx}

\title{Investigating the Relationship Between User Specialization and Toxicity on Reddit: A Sentiment Analysis Approach}

\author{Abi Oppenheim\textsuperscript{1}\and
Federico Albanese\textsuperscript{2,3}\and
Esteban Feuerstein\textsuperscript{1,3} \\ 
\texttt{\{aoppenheim, falbanese, efeuerst\}@dc.uba.ar} \\ \\
\begin{minipage}{1\textwidth}
\centering
{\fontsize{10}{11}\selectfont
{\textsuperscript{1}Departamento de Computación, Facultad de Ciencias Exactas y Naturales,
\\Universidad de Buenos Aires, Buenos Aires, Argentina.} \\[0.5em] 
\textsuperscript{2}Instituto de Cálculo, CONICET Universidad de Buenos Aires, Buenos Aires, Argentina. \\[0.5em] 
\textsuperscript{3}Instituto de Ciencias de la Computación, CONICET Universidad de Buenos Aires, Buenos Aires, Argentina.
}
\end{minipage}
}

\begin{document}
\maketitle
\begin{abstract}
Online platforms host users with diverse interests, ranging from those with focused interests to those engaging in a wide range of topics. This study investigates the behavioral differences between these user types on Reddit, focusing on the level of toxicity in their posts and associated sentiment scores across nine emotional categories. By employing community embeddings to represent users in a high-dimensional space, we measure activity diversity using the GS score. The analysis utilizes a dataset of 16,291,992 posts from 4,926,237 users spanning 2019 to 2021, assessing toxicity and sentiment scores for each post. Results indicate that subreddits characterized by users with specialized interests exhibit heightened toxic behavior compared to those with diverse interest users. Additionally, subreddits populated by users with focused interests display elevated sentiments of sadness, annoyance, and disappointment, while those inhabited by diverse interest users demonstrate increased expressions of curiosity, admiration, and love. \\These insights contribute to understanding user behavior on online platforms and inform strategies for fostering healthier online communities.
\end{abstract}

\section{Introduction}
Social media platforms have become breeding grounds for toxic behaviors, hate speech, and online harassment, posing significant challenges to digital communities. A recent report~\cite{thomas2021sok} underscores the widespread prevalence of online threats, with nearly half (48\%) of individuals worldwide reporting encounters with such menacing behavior. Marginalized communities, in particular, bear a disproportionate burden of online harassment~\cite{golbeck2018online, massanari2017gamergate}, leading to detrimental effects on mental health and civil discourse.
Various studies have explored methods for detecting and analyzing toxic texts and emotions, including sentiment analysis on COVID-19 vaccine-related discussions~\cite{melton2021public}, examining toxicity in political contexts~\cite{albanese2023aprendizaje}, analyzing toxic language usage on Facebook in relation to political interest~\cite{kim2021distorting}, and studying toxicity in conversations prompted by tweets from news outlets and political candidates\cite{saveski2021structure}.
Building upon this research, our work integrates these strands with the methodology proposed by Waller and Anderson~\cite{waller2019generalists}. Their study investigates user behavior by distinguishing between generalist and specialist tendencies and utilizes community embeddings to represent online communities. They employ the community2vec algorithm~\cite{ref_community2vec} to assign vectors to communities, facilitating the quantification of similarity between communities based on cosine similarity in the vector space. Additionally, they introduce the Generalist-Specialist score (GS-score) to measure a user's activity diversity, where specialists have a low GS-score, contributing to a tight cluster of communities, while generalists have a high GS-score, contributing to diverse, distant communities.

However, up to our knowledge, the relationship between user activity diversity and the level of toxicity in their online interactions remains unexplored. In this study, we hypothesize that communities comprising users with a broader array of interests, are less likely to have harmful behavior compared to communities dominated by those who concentrate on a limited range of topics.
To test our hypothesis, we analyze a dataset of 16,291,992 posts from 4,926,237 users on Reddit, spanning the period from 2019 to 2021. By employing community embeddings to represent users in a high-dimensional space, we measure activity diversity using the GS score. We assess the degree of toxicity and sentiment scores across 9 emotional categories for each post. 
Our findings reveal two key observations. 
\begin{itemize}
    \item \textbf{Diversity of Interest and Toxic Behavior}. Our investigation shows that subreddits cultivated by individuals with specialized interests experience more toxic behavior than those curated by individuals with diverse interests.
    
    \item \textbf{Emotions Across Communities Types}. Subreddits shaped by specific interest users show more sadness, annoyance, and disappointment, while those by diverse interest users display higher levels of admiration, amusement, and affection.
\end{itemize}
These findings underscore a compelling link between the diversity of user interests and a notable decrease in toxic behavior observed on social media platforms.


\section{Methodology}
\subsection{Dataset}
We utilized the Pushshift API, a platform for collecting, analyzing, and archiving social media data~\cite{ref_pushshift}. The Pushshift dataset encompasses submissions and comments posted on Reddit since June 2005, offering extensive query limits and user-friendly access. To ensure scalability and representativeness, we randomly selected 15,000 submissions from each of the top 5,000 most popular subreddits, evenly distributed across 2019 to 2021. Our dataset comprises 16,291,992 posts from 4,926,237 users.

\subsection{Generalist-Specialist Score (GS-score)}
The GS-score quantifies user activity diversity, distinguishing between specialized and generalized engagement patterns~\cite{waller2019generalists}. Specialized users concentrate activity within specific areas of the community space, while generalist users show broader dispersion.

To analyze user interactions across diverse communities, we first generate community embeddings using the community2vec approach~\cite{waller2021quantifying, ref_community2vec}. By treating communities as ``words'' and users submitting in them as ``contexts'', communities were embedded into a high-dimensional vector space. The proximity of two communities in this space is determined by the frequency of users posting in both communities. We trained the skip-gram model with negative sampling using all pairs $(c_i, u_j)$ of users $u_j$ submitting in community $c_i$, to generate an initial community embedding.

Let user $u_i$ make $w_j$ contributions to community $c_j$, and let $\vec{c_j}$ denote the normalized vector representation of $c_j$ in our community embedding. The center of mass of $u_i$ is defined as $\vec{\mu_i} = \sum(w_j * \vec{c_j})$. The diversity of $u_i$'s activity, or GS-score, is the weighted average cosine similarity between $u_i$'s communities and its center of mass:

$$GS(u_i) = \frac{1}{\sum w_j} \sum_j w_j \frac{\vec{c_j} \cdot \vec{\mu_i}}{\|\vec{\mu_i}\|}$$
The GS-score ranges from $-1$ to $1$, where $-1$ indicates extreme generalization and $1$ indicates extreme specialization. For detailed insights, refer to Waller and Anderson~\cite{waller2019generalists}.
Once user activity diversity is computed, we extend this analysis to understand community activity diversity. The GS-score $GS(c_i)$ of community $c_i$ is the weighted average across its users: $GS(c_i) = \frac{1}{N} \sum_j w_j \cdot GS(u_j)$, where $u_j$ contributes $w_j$ times to $c_i$ and $N = \sum_j w_j$ total contributions. Prioritizing community metrics enhances robustness and comprehensive evaluation of submission sets within communities.

\subsection{Emotion and Toxicity Analysis}
For emotion classification, we utilized the pre-trained Language Model (LLM) "SamLowe/roberta-base-go\_emotions"~\cite{demszky2020goemotions}, fine-tuned on the GoEmotions dataset~\cite{ref_roberta}, which consists of 58,000 Reddit comments annotated across 27 emotional categories, including Neutral. This model's training on Reddit-specific data makes it well-suited for our analysis and has been utilized in various scientific studies~\cite{kocon2023chatgpt, davani2022dealing, vu2021spot}.
Considering the vast range of emotions, we focused on specific subsets relevant to our hypothesis, including admiration, amusement, anger, annoyance, disappointment, gratitude, joy, love, and sadness. This model was applied to over 200 pertinent subreddits within our corpus.

To evaluate toxicity, we employed the compact Detoxify Model developed by Unitary AI~\cite{ref_detoxify}, trained on a substantial dataset of Wikipedia comments annotated for various forms of toxic behavior~\cite{ref_albert}. This model outputs probabilities indicating the likelihood of input text containing language associated with different toxic categories. Over 1100 relevant subreddits were analyzed using this approach.

\section{Results}

\subsection{Subreddit Distribution Analysis}

\begin{figure}[h]
  \includegraphics[width=\columnwidth]{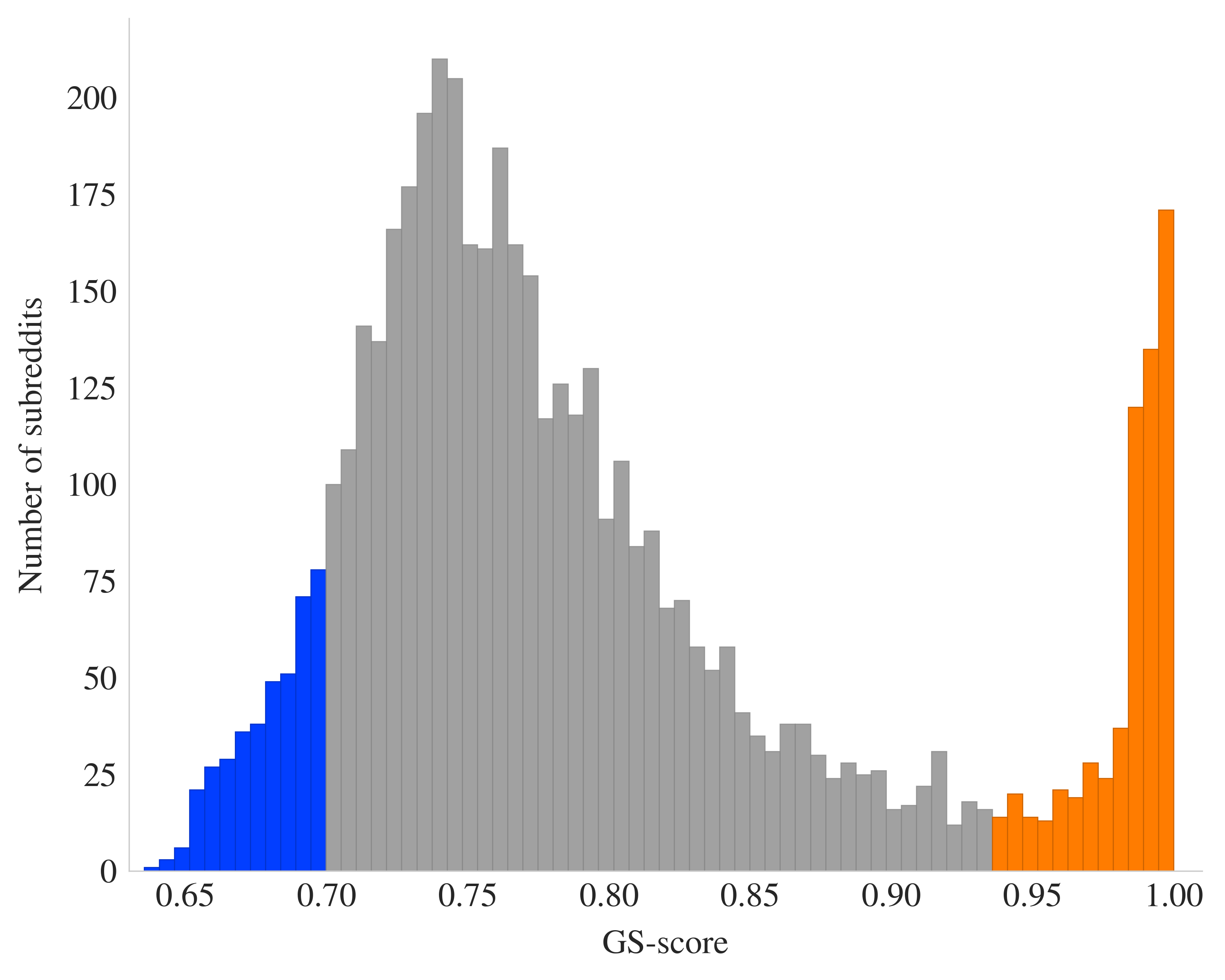}
  \caption{Distribution of GS-scores among subreddits}
  \label{fig:fig1_GSScore_histogram}
\end{figure}

To examine user behavior across subreddit types, we categorized the subreddits into two groups: the bottom 10th percentile of GS-scores, representing specialized-interest user communities, and the upper 10th percentile, representing diverse-interest user communities, each comprising 20 subreddits. To gain insights into the dataset's distribution, Figure \ref{fig:fig1_GSScore_histogram} presents a histogram illustrating the distribution of GS-scores among the subreddits. 
The blue bars represent subreddits within the lower 10th percentile of GS-scores, while the orange bars represent those within the upper 10th percentile. Notably, the GS-scores predominantly cluster within the range of 0.6 to 1, aligning with prior findings~\cite{waller2019generalists}.

\subsection{Emotion and Toxicity Analysis}
\begin{figure}[h]
    \centering
    \includegraphics[width=1\linewidth]{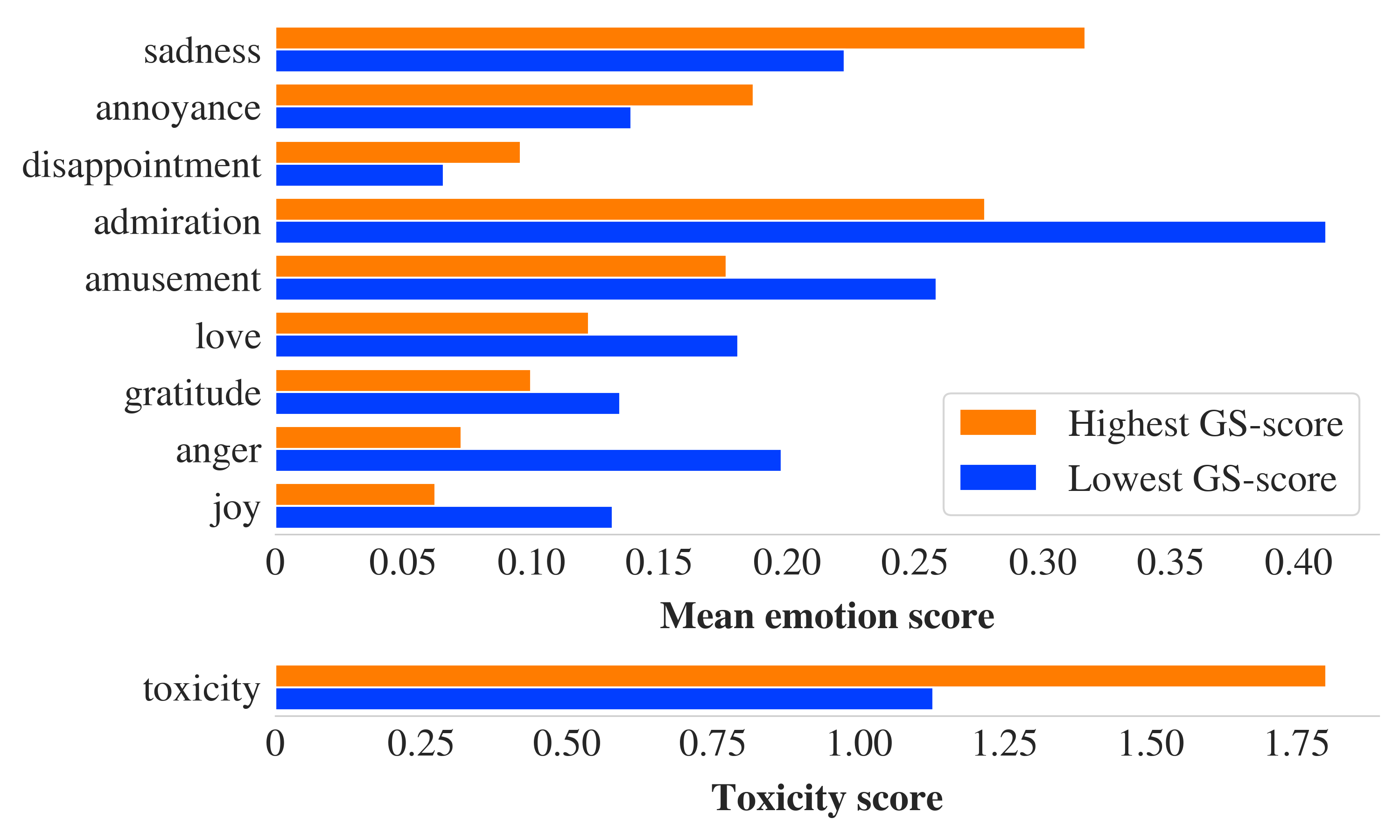}
    \caption{Emotion and toxic scores comparison between subreddits with highest and lowest GS-score}
    \label{fig:fig3_Specialist_vs_Generalist}
\end{figure}

For each group, we calculated average emotion and toxicity scores by determining the proportion of users expressing specific emotions relative to the total number of users. To explore variations in emotional expression, we employed the Mann-Whitney U test, suitable for comparing two independent groups when normality assumptions are not met.
Our findings indicate significant differences between subreddits frequented by users with diverse and specialized interests, as evidenced by p-values smaller than 0.05. Subreddits with the highest GS-scores exhibited higher toxicity scores compared to those with the lowest GS-scores ($\boldsymbol{r=.44}$). Conversely, subreddits with the lowest GS-scores demonstrated higher scores for emotions such as joy ($\boldsymbol{r=.48}$), anger ($\boldsymbol{r=.56}$), gratitude ($\boldsymbol{r=.44}$), love ($\boldsymbol{r=.53}$), amusement ($\boldsymbol{r=.58}$), and admiration ($\boldsymbol{r=.4}$).Notably, all statistically significant results align with our hypothesis. However, disappointment ($p=.394, r=0.16$) and sadness ($p=.968, r=.01$) did not show significant differences between the two types of subreddits. Figure \ref{fig:fig3_Specialist_vs_Generalist} compares emotion and toxicity scores between subreddit types.

While the lowest GS-score subreddits demonstrated a higher score for anger compared to the other subreddit type, it is essential to differentiate between ``angry'' and ``toxic'' content in text analysis. The distinction between anger and toxicity lies in their respective expressions: while anger typically manifests as frustration towards situations, objects, or groups, toxicity often involves directing abusive language or behavior towards individuals.

\section{Conclusions}
Our study delved into the connection between emotion, toxicity, and activity diversity on Reddit. We found that subreddits characterized by users with diverse interests tend to have lower toxicity and higher positive emotion scores, including joy, gratitude, love, and admiration. Conversely, subreddits shaped by users with more specific interests exhibit higher toxicity and increased negative emotions like sadness, annoyance, and disappointment. These findings underscore the significance of activity diversity in understanding user behavior and its impact on online community dynamics. Our research enhances comprehension of the factors influencing user conduct and offers insights for fostering healthier online interactions. Future studies could investigate causal links between activity diversity, emotion, and toxicity, as well as explore interventions to encourage positive engagement and mitigate toxicity on online platforms.


\bibliography{acl}

\end{document}